\documentclass[final,5p,times,twocolumn]{elsarticle}

\usepackage{amssymb,amssymb,amsfonts}
\usepackage{multirow}
\usepackage[caption=false,font=footnotesize]{subfig}
\usepackage{caption}
\usepackage{booktabs}
\usepackage{amsthm}
\usepackage{amsmath}
\usepackage{graphicx}
\usepackage{makecell}
\usepackage{textcomp}
\usepackage[ruled,linesnumbered]{algorithm2e}
\usepackage{color}
\usepackage{colortbl}
\usepackage[font=small,labelfont=bf,labelsep=none]{caption}
\usepackage{xcolor}
\usepackage[draft]{changes}
\definecolor{mygray}{rgb}{0.85,0.85,0.85}
\definecolor{myblue}{rgb}{0.78,0.99,1}

\journal{Elsevier}

\begin{document}

\begin{frontmatter}



\title{ARFT-Transformer: Modeling Metric Dependencies for Cross-Project Aging-Related Bug Prediction}


\author[1]{Shuning Ge}
\ead{shuningge@cnu.edu.cn}
\author[1]{Fangyun Qin\corref{cor1}}
\ead{fyqin@cnu.edu.cn}
\author[2]{Xiaohui Wan}
\ead{wanxh@aircas.ac.cn}
\author[3]{Yang Liu}
\ead{yangliu2@bjtu.edu.cn}
\author[4]{Qian Dai}
\ead{qiandai2024@gmail.com}
\author[5]{Zheng Zheng}
\ead{zhengz@buaa.edu.cn}

\cortext[cor1]{Corresponding author}
\address[1]{College of Information Engineering, Capital Normal University, Beijing, China}
\address[2]{Suzhou Aerospace Information Research Institute, Suzhou, China}
\address[3]{School of Mechanical, Electronic and Control Engineering, Beijing Jiaotong University, Beijing, China}
\address[4]{Beijing Institute of Computer Technology and Applications, Beijing, China}
\address[5]{School of Automation Science and Electrical Engineering, Beihang University, Beijing, China}

\begin{abstract}
Software systems that run for long periods often suffer from software aging, which is typically caused by Aging-Related Bugs (ARBs). To mitigate the risk of ARBs early in the development phase, ARB prediction has been introduced into software aging research. However, due to the difficulty of collecting ARBs, within-project ARB prediction faces the challenge of data scarcity, leading to the proposal of cross-project ARB prediction. This task faces two major challenges: 1) domain adaptation issue caused by distribution difference between source and target projects; and 2) severe class imbalance between ARB-prone and ARB-free samples. Although various methods have been proposed for cross-project ARB prediction, existing approaches treat the input metrics independently and often neglect the rich inter-metric dependencies, which can lead to overlapping information and misjudgment of metric importance, potentially affecting the model's performance. Moreover, they typically use cross-entropy as the loss function during training, which cannot distinguish the difficulty of sample classification. To overcome these limitations, we propose ARFT-Transformer, a transformer-based cross-project ARB prediction framework that introduces a metric-level multi-head attention mechanism to capture metric interactions and incorporates Focal Loss function to effectively handle class imbalance. Experiments conducted on three large-scale open-source projects demonstrate that ARFT-Transformer on average outperforms state-of-the-art cross-project ARB prediction methods in both single-source and multi-source cases, achieving up to a 29.54\% and 19.92\% improvement in Balance metric.

\end{abstract}




\end{frontmatter}


\section{Introduction}
\label{introduction}
Software systems are typically built with long-term operation in mind, aiming to provide stable and reliable service over time. However, such long-running software often faces the challenge of software aging \cite{huang1995software}. Typically, software rejuvenation is regarded as an effective strategy to mitigate software aging's adverse effects \cite{huang1995software,dohi2020handbook,castelli2001proactive,trivedi2007software}, but it cannot fundamentally eliminate Aging-Related Bugs (ARBs) which is the reason leading to software aging. ARB prediction, by extracting metrics from software in its early development stages and using them to build the predictor with machine learning techniques, enables the early identification of ARB-prone areas and helps ARB removal, addressing the issue fundamentally \cite{chouhan2021generative,zhang2024sgt,tian2025towards,qin2018studying,wan2019supervised, jia2023software}. Since Cotroneo et al. pioneered research on ARB prediction in 2013 \cite{cotroneo2013predicting}, ARB prediction has emerged as a growing research hotspot.

However, it is a difficult task to conduct traditional within-project ARB prediction (i.e., building the prediction model with training data from the target project). Firstly, due to its long-term error accumulation leading to aging-related failure, it is a challenging task to collect relevant datasets \cite{cotroneo2014survey,grottke2010empirical,grottke2008fundamentals}. Secondly, ARBs constitute a small proportion of total bugs \cite{cotroneo2013fault,grottke2010empirical,grottke2008fundamentals}, requiring the analysis of vast bugs to obtain sufficient aging-related training samples. Thirdly, for a project in its initial development or without historical archived bug data, it is even more hard to conduct the prediction. To tackle the aforementioned problems, researchers have proposed cross-project ARB prediction by accomplishing the ARB prediction task in the target project with training data from a different but related project.


Nevertheless, cross-project ARB prediction faces two major challenges. For one thing, the distribution difference between source and target project poses threats to the adaptability of machine learning classifiers with same distribution assumption between training and testing set. For another thing, the severe class imbalance existing between ARB-prone and ARB-free samples poses a negative impact on the performance of classifiers. To overcome these two challenges, researchers have proposed several approaches by combining transfer learning and class imbalance mitigation techniques, such as TLAP with Transfer Component Analysis (TCA) and random oversampling \cite{qin2015cross,qin2018studying}, SRLA with autoencoder and random oversampling (ROS) \cite{wan2019supervised}, JDA-ISDA with Joint Distribution Adaptation (JDA) and Improved Sub-class Discriminant Analysis (ISDA) \cite{xu2020cross}, JPKS with Joint Probability Domain Adaptation (JPDA) and k-means SMOTE (KS) \cite{li2021cross}, KDK with Kernel Principal Component Analysis (KPCA), Double Marginalized Denoising Autoencoder (DMDA), and K-means Clustering Cleaning Ensemble (KCE) \cite{xie2023cross}, BISP with balanced distribution adaptation (BDA) and the self-paced ensemble under-sampling (SPE) \cite{xu2025cross}.

While existing cross-project ARB prediction approaches achieved a large improvement on the prediction performance, all of them treat the input metrics in a sample independently. That is, without considering the possible correlations among metrics. For example, a high $\mathit{CountLineCode}$ metric value usually accompanies with a high $\mathit{SumCyclomatic}$ value. This correlation ignorance may lead to overlapping information among highly related metrics and underestimating or overestimating the importance of certain metrics, which may negatively affect the prediction model's performance \cite{jiarpakdee2018impact, tashtoush2014correlation, mamun2017correlations}. Table~\ref{tab:relationships} lists the correlation among 52 metrics (1326 metric pairs) in the benchmark of ARB prediction dataset with Spearman's rank correlation test \cite{spearman1961proof}. From the table, we can find that 91.86\%, 90.57\%, and 83.94\% of metric pairs in the three datasets exhibit significant correlations ($|\rho| > 0.3$, $p < 0.05$), indicating that most metrics are not independent each other.

In addition, to mitigate the class imbalance during model training, previous research typically relied on basic sampling strategies such as Random Oversampling \cite{zhang2024sgt}, SMOTE \cite{kumar2017aging}, k-means SMOTE \cite{xie2023cross}, Weakly Supervised Oversampling \cite{zhou2022software}, and Random Undersampling \cite{xu2020cross} to balance the number of samples between ARB-prone and ARB-free classes. While these techniques helped alleviate class imbalance, it may not be sufficient on their own as they can only increase the classifier's focus on the ARB-prone samples. The used standard cross-entropy loss assigned the same loss weight to all samples of the same class, without distinguishing samples' hard-to-classify property \cite{lin2017focal,shrivastava2016training,wang2019symmetric}. As a result, within the same class, the numerous easy-to-classify samples contribute disproportionately to the overall loss. Due to the large number of easily classified ARB-free samples in ARB prediction, their accumulated loss tends to dominate the training process, thereby suppressing the learning of the sparse ARB-prone samples.

In this paper, we propose ARFT-Transformer, a cross-project ARB prediction approach based on FT-Transformer \cite{gorishniy2021revisiting} and Focal loss \cite{lin2017focal} to consider both metric correlation and sample classification hardness. Specifically, ARFT-Transformer introduces a metric-level multi-head attention mechanism based on FT-Transformer \cite{gorishniy2021revisiting} to dynamically capture the relational characteristics among metrics. Each metric is transformed into a unique representation vector and Maximum Mean Discrepancy (MMD) loss \cite{long2015learning} is used to align the learned representation distributions between source and target datasets. To tackle the adverse effect of class imbalance during model training, while using ROS method to randomly increase the number of ARB-prone samples, ARFT-Transformer designs a Focal Loss function \cite{lin2017focal} to handle hard-to-classify samples.

To evaluate the effectiveness of the proposed approach, we conduct both single-source and multi-source experiments with three commonly and widely used datasets in the area of ARB prediction, where single-source setting refers to using one project as the source project to train the prediction model and multi-source setting denotes combining multiple projects together (which may bring more information about ARB) as source set to accomplish the training task. We compare ARFT-Transformer's prediction performance in each setting with several state-of-the-art approaches. Besides, to assess the contribution of multi-head attention \cite{vaswani2017attention} and Focal Loss \cite{lin2017focal} which acts as two key components of ARFT-Transformer, ablation studies are designed. In addition, we discuss the impact of hyperparameter selection during model training.

The contribution of this paper is as follows:
\begin{itemize}
\item We propose an approach named ARFT-Transformer, a neural network framework enhanced with multi-head attention to model metric relationships, for cross-project ARB prediction tasks. To the best of our knowledge, this is the first approach modeling correlations among code metrics in ARB prediction.
\item We introduce Focal Loss along with ROS to not only focus more on ARB-prone samples but also to hard-to-classify samples, thereby mitigating the class imbalance problem during model training.
\item We perform both single-source and multi-source cross-project ARB prediction across three widely-used projects in ARB prediction to validate its effectiveness. The results indicate that ARFT-Transformer exhibits promising prediction capabilities in both prediction cases.
\end{itemize}

The rest of the paper is organized as follows. Section \ref{sec:background} introduces the background. Section \ref{sec:approach} presents the proposed model, Section \ref{sec:preparation} describes experimental setup and Section \ref{sec:results} reports experiments results. Section \ref{sec:discussion} discusses the impact of hyperparameter selection on the model's performance. Section \ref{sec:threats_to_validity} explores the threats to validity in the paper and Section \ref{sec:relatedwork} presents the related work. Section \ref{sec:conclusion} concludes.

\begin{table}[!t]
\renewcommand{\arraystretch}{1.2}
\caption{\textsc{Significant Correlated Metric Pairs in Each Project}}
\vspace{-3mm}
\centering
\begin{tabular}{>{\raggedright\arraybackslash}m{18mm} 
                >{\centering\arraybackslash}m{15mm} 
                >{\centering\arraybackslash}m{18mm} 
                >{\centering\arraybackslash}m{18mm}}
\hline
\textbf{Projects} & \textbf{Total Metrics  Pairs} & \textbf{Correlated Metric Pairs} & \textbf{Correlated Metric Pairs\%} \\
\hline
Linux  & 1326 & 1218 & 91.86\% \\
MySQL  & 1326 & 1201 & 90.57\% \\
HTTPD  & 1326 & 1113 & 83.94\% \\
\hline
\end{tabular}
\label{tab:relationships}
\end{table}

\section{Background}
\label{sec:background}
In this section, we provide background information on software bug prediction.

Software bug prediction refers to using information such as code complexity, software development history, and developers to predict which modules may contain bugs \cite{zhao2023systematic,wang2018deep,hosseini2017systematic,nevendra2022survey}. Depending on the studied granularity, the module here can represent software package, source code file, class, function or code change.

\begin{figure}[!t]
    \centering
\includegraphics[width=\columnwidth]{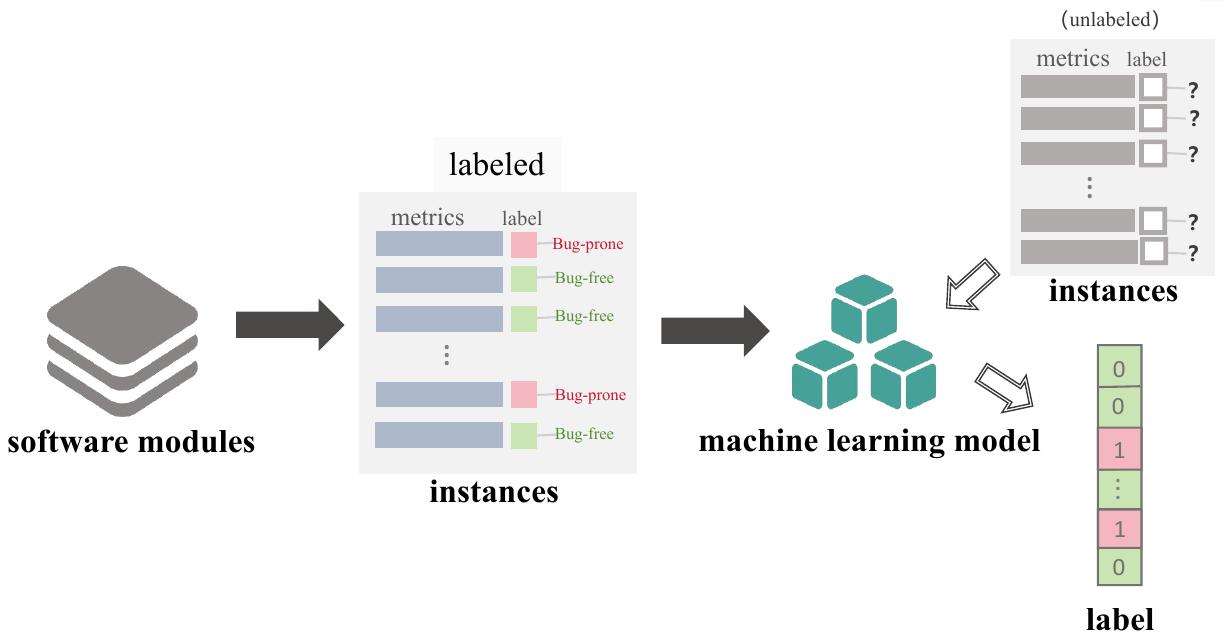}
    \caption{The process of software bug prediction.}
    \label{fig:software_bug_p}
\end{figure}

The process of Software bug prediction is shown as Figure \ref{fig:software_bug_p}. To construct a software bug prediction model, researchers typically begin by collecting historical project data and labeling software modules as either bug-prone or bug-free. Subsequently, various software metrics that reflect different dimensions of complexity, such as lines of code and control flow complexity, are extracted for each module \cite{bibi2006software,sun2012using,shepperd2014researcher,nevendra2022survey,giray2023use}. These metrics, together with the module labels, form labeled instances (also referred to as samples), which collectively constitute a labeled dataset. In this dataset, the extracted metrics serve as independent variables, while the module label acts as the dependent variable. Based on this structured data, prediction models—often built using machine learning techniques—are trained to identify patterns and correlations between code metrics and the likelihood of bugs. Once trained, the model can be applied to new code modules (testing data) by inputting their metrics to output a classification flagging buggy risk. This enables development teams to efficiently prioritize testing and code review efforts, focusing resources on components most susceptible to bugs, thereby enhancing software quality and reducing maintenance costs.

Commonly used evaluation measures to assess the performance of bug prediction model are derived from the confusion matrix, which comprises four fundamental elements: True Positive (TP) denotes bug-prone modules correctly identified as buggy; False Negative (FN) refers to buggy modules that are incorrectly classified as bug-free; False Positive (FP) indicates bug-free modules wrongly labeled as buggy; and True Negative (TN) represents bug-free modules accurately classified as bug-free. Based on these values, a suite of performance measures can be calculated. Further details are provided in Section \ref{subsec: model_evaluation_criteria}.

\section{Approach}
\label{sec:approach}
In this section, we introduce the proposed ARFT-Transformer approach for cross-project ARB prediction, aiming at addressing three core challenges: effective ARB feature extraction, cross-project domain adaptation, and class imbalance during prediction model training. We first present the overall framework of ARFT-Transformer and then introduce each part in detail.

\subsection{Overall framework}
\begin{figure*}[htbp]
    \centering
    \includegraphics[width=0.9\textwidth]{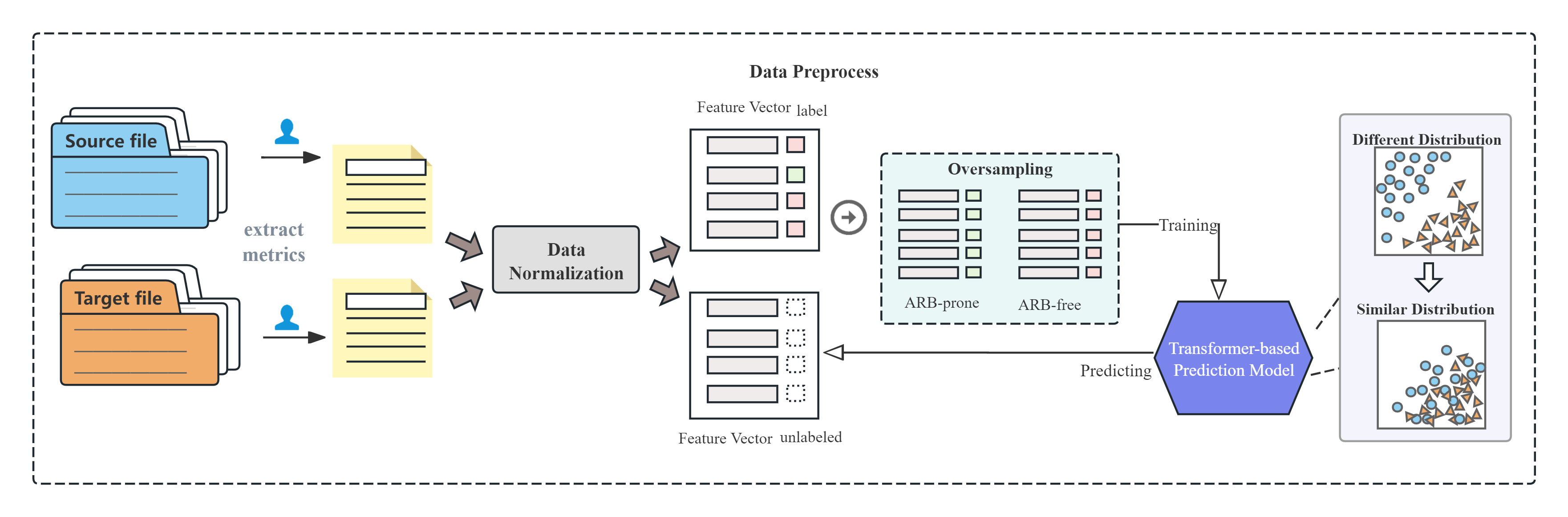}
    \vspace{-2mm}
    \caption{The overall framework of ARFT-Transformer.}
    \label{fig:overall}
    \vspace{-4mm}
\end{figure*}

The overall workflow of ARFT-Transformer is illustrated in Figure~\ref{fig:overall}. In the experiment, the static code metrics used are extracted from individual files, with each file represented as a feature vector. The process is roughly divided into three steps. First, data preprocessing including normalization and oversampling. Normalization is applied to both the source and target samples, while oversampling is applied only to the source samples. Then, the processed data are fed into Transformer-based model, where they pass through a tokenizer, multi-head attention mechanism, and MMD-based distribution alignment. Finally, we introduce the model training. 

\begin{figure*}[!t]
    \centering
    \includegraphics[width=0.9\textwidth]{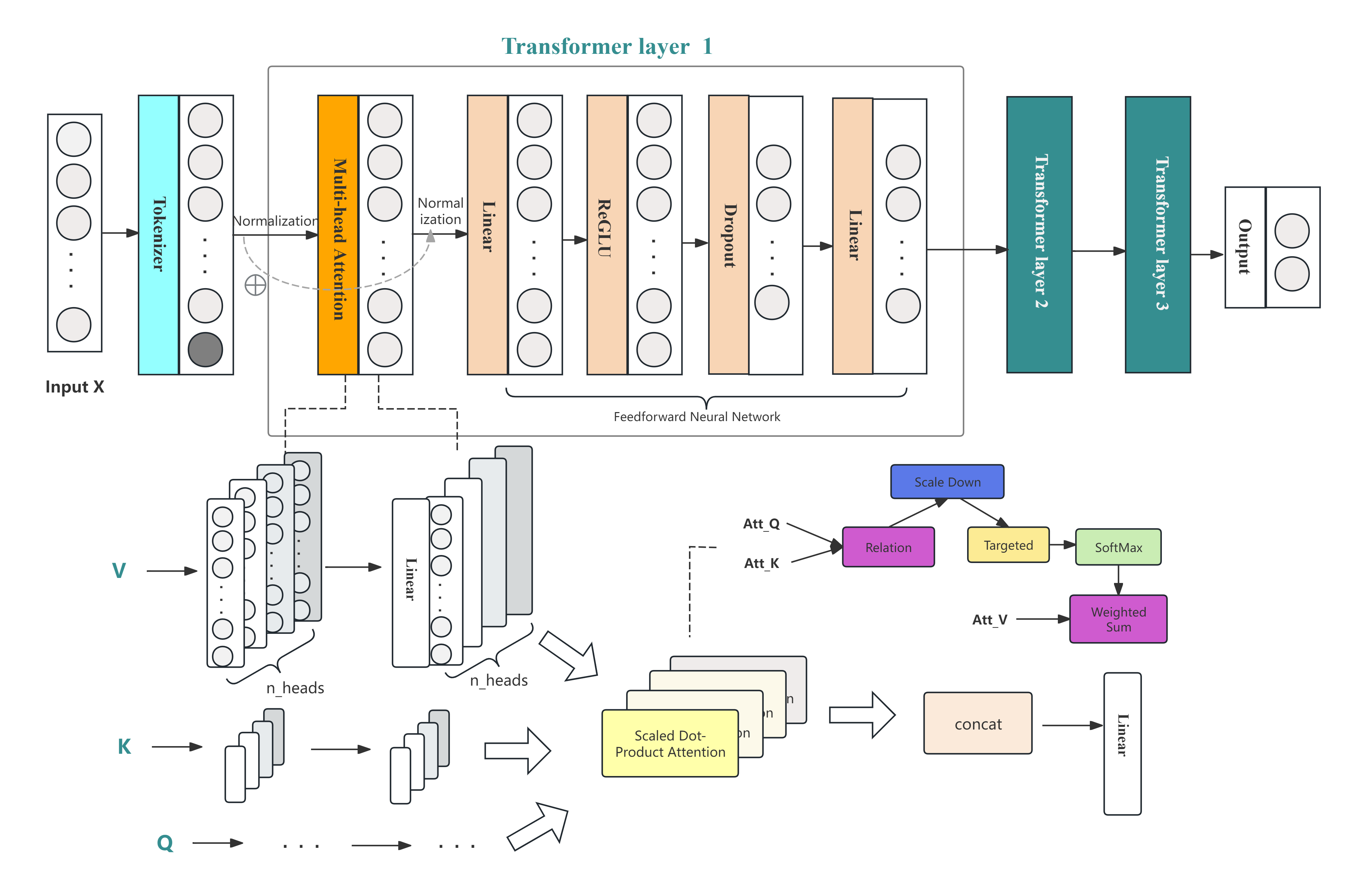}
    \vspace{-2mm}
    \caption{Structure of FT-Transformer.}
    \label{fig:model}
    \vspace{-2mm}
\end{figure*}

\subsection{Data Preprocessing}\label{AA}
At the outset of the experiment, we preprocess the raw datasets with Z-score normalization \cite{mining2006data} to scale the data to a uniform range. Specifically, by combining source and target domain data, we first compute the global mean and global standard deviation, and then apply these two statistics to standardize all data, including the source domain training data and the target domain testing data. By jointly normalizing the data from the source and target domains, their distributions become more similar \cite{apicella2023effects}.

The source and target data are represented as
\begin{equation}
\begin{split}
\{X_s,Y_s\}&\!=\!\left \{ \left ( x_{11},\ldots,x_{1p},y_{1} \right ),\ldots,\left ( x_{N_s1},\ldots,x_{N_sp},y_{N_s} \right ) \right \} \\
\{X_t\}&\!=\!\left \{ \left ( x_{11},\ldots,x_{1p} \right ),\ldots,\left ( x_{N_t1},\ldots,x_{N_tp} \right ) \right \}
\end{split}
\end{equation}
where $X_s$ and $X_t$ refer to the feature matrix in source and target project respectively, $Y_s$ refers to the sample labels in the source project. $N_s$ and $N_t$ denote the number of samples in the source and target domain respectively, $p$ denotes the number of features (metrics) in each sample. Then the concatenated matrix can be represented as:
\begin{equation}
X_{all}= \left ( X_s,X_t \right ).\label{eq}
\end{equation}
In the domain adaptation task, we adopt a global standardization approach, calculating the global mean $\mu_g$ and global standard deviation $\sigma_g$:
\begin{equation}
\mu _g= \frac{1}{N_S+N_t}\sum_{i=1}^{N_s+N_t} x_i,\label{eq}
\end{equation}
\begin{equation}
\sigma _g=\sqrt{\frac{1}{N_S+N_t}\sum_{i=1}^{N_s+N_t}\left ( x_i-\mu_g\right ) ^2  } ,\label{eq}
\end{equation}
Here, $x_i$ represents the feature vector of the $i$-th sample.

Based on this, the normalized feature is given by:
\begin{equation}
{X}_{norm}\left ( i,j \right ) =\frac{{X}_{all}\left ( i,j \right ) -\mu _g}{\sigma_g+\epsilon } , \label{eq}
\end{equation}
where $\epsilon$ is a small constant introduced to prevent division by zero, ensuring numerical stability. We apply this global normalization method to project all data into a unified distribution space, ensuring that the statistical properties of the source and target domain features are aligned.

To overcome the class imbalance, we also apply Random Oversampling (ROS) to augment the source project dataset before model training, addressing the severe imbalance between ARB-prone and ARB-free samples. ROS randomly duplicates minority class samples to match the number of majority class samples, thereby mitigating the model’s bias toward the majority class and enhancing its learning capability for the minority class. However, in the target project, we refrain from using any class imbalance techniques to simulate the distribution characteristics of imbalanced datasets in real-world scenarios.

\subsection{FT-Transformer}
In recent years, the Transformer architecture has gained widespread adoption in natural language processing and computer vision \cite{dosovitskiy2020image,khan2022transformers,han2022survey,rogers2021primer}. Building on this foundation, FT-Transformer was proposed as an effective adaptation of the Transformer model for tabular data. Its model structure is shown in Figure~\ref{fig:model}.

FT-Transformer introduces a specialized [CLS] token to aggregate information across all features and employs a multi-layer Transformer structure to capture complex interactions among features. The workflow of the model can be summarized as follows. Let $X_{\text{p}}$ denotes the feature matrix without [CLS] token, where $p$ represents the number of features. After expansion, its dimensionality becomes $p+1$, with the additional dimension reserved for the [CLS] token, as shown below
\begin{equation}
X = \text{concat}([\mathbf{1}_{N \times 1}, X_{\text{p}}], \text{axis} = 1) \in \mathbb{R}^{N \times (p + 1)} .\label{eq}
\end{equation}
Each input sample is mapped to an embedding space through a tokenizer that linearly transforms each feature into a token vector containing richer and more complex information, resulting in a matrix denoted as $X_{\text{embed}}$:
\begin{equation}
X_{\text{embed}} = X W_{\text{embed}} + b_{\text{embed}} \in \mathbb{R}^{N \times (p + 1) \times d_{\text{token}}} \label{eq}
,\end{equation}
where $d_{\text{token}}$ represents the dimension of each feature vector.

Then $X_{\text{embed}}$ is concatenated to form a matrix that includes the [CLS] token, denoted as  
\begin{equation}
X_0 = X_{\text{embed}}^{[\text{CLS}]} \in \mathbb{R}^{N \times (p+1) \times d_{\text{token}}}
.\label{eq}
\end{equation}
Here, the [CLS] token is a learnable vector that aggregates information, and the additional position for it has been reserved before. After processing through Transformer layer, the final representation is obtained as
\begin{equation}
X_l = F_l(X_{l-1}) \quad \text{for} \quad l = 1, \dots, L.
\end{equation}
Here, $X_l$ represents the output after processing through the $l$-th Transformer layer, $F_l$ is the set of functions for the $l$-th layer, and $L$ denotes the layer number. In this paper, we design three layers of Transformer.

Each Transformer layer consists of two parts: the Multi-head Attention and the Feed-Forward Neural Network. Here we take the first Transformer layer as an example. In the attention part, input is processed by multiple heads in parallel, and each head follows the standard formulation:
\begin{equation}
\mathrm{head}_m \!=\! \mathrm{Attention}(Q_m, K_m, V_m) \!=\! \mathrm{softmax}\left( \frac{Q_m K_m^T}{\sqrt{d_k}} \right) V_m, \label{eq}
\end{equation}
where $m$ denotes the number of heads and $d_k$ represents the dimension of $K$. In this paper, we set the number of heads to 8 following the research by Vaswani et al. \cite{vaswani2017attention}. Each head computes attention over the input matrix $X$, including the [CLS] token, using query $Q$, key $K$, and value $V$, as follows:
\begin{equation}
Q_m = X W_{q,m}, \quad K_m = X W_{k,m}, \quad V_m = X W_{v,m}.\label{eq}  \end{equation}
Let $W^O$ denotes the output  projection matrix. The final output of the multi-head attention which contains the feature interaction information is given by:
\begin{equation}
\mathrm{MultiHead}(Q, K, V) = \mathrm{Concat}(\mathrm{head}_1, \dots, \mathrm{head}_m) W^O  .
\label{eq:multihead}
\end{equation}

Finally, the output is passed through a Feed-Forward Neural Network for nonlinear transformation, which includes ReGLU as the activation function and a dropout layer to prevent overfitting by randomly deactivating neurons. The prediction is denoted as:
\begin{equation}
\hat{y} = \text{Linear}\left(\text{LayerNorm}\left(\text{Transformer}_L(X)^{[\text{CLS}]}\right)\right),
\end{equation}
where ${\left ( X\right) ^\text{[CLS]}}$ denotes the output at the [CLS] token position from the final Transformer layer, which is used for prediction. 

\subsection{Model training}

To address class imbalance and domain distribution discrepancy, we design a composite loss function that combines classification loss with feature alignment loss, as shown below
\begin{equation}
\mathcal{L} = \mathrm{FL}(p_t) + \lambda \cdot \mathrm{MMD}(X_s, X_t),\label{eq:total programmed_loss}
\end{equation}
where FL refers to Focal Loss \cite{lin2017focal} and MMD represents Maximum Mean Discrepancy. By introducing the weighting factor $\lambda$, the model dynamically adjusts its focus on the feature alignment loss during iterations. In the early stage of training, the model prioritizes the classification task, gradually incorporating feature alignment in later stage to avoid interfering with classification learning at the outset. 

Focal Loss is built upon cross-entropy loss by Lin et. al \cite{lin2017focal}. It can dynamically adjust sample weighting by introducing a modulating factor, thereby increasing focus on hard-to-classify samples:
\begin{equation}
\mathrm{FL}(p_t) = -\alpha (1 - p_t)^\gamma \log(p_t).  \label{eq:focal_loss}
\end{equation}
Here, $p_t$ denotes the model’s predicted probability for the true class of the sample, $\gamma$ represents the modulating factor that controls the focus on hard-to-classify samples, and $\alpha$ is the class weighting factor used to further balance the contributions of ARB-prone and ARB-free classes, which is set to 1 in our experiments. Notably, when $\gamma$ is set to 0, Focal loss is equivalent to the standard weighted cross-entropy loss. In this paper, $\gamma$ is set to 2 following the setting of Lin et al. \cite{lin2017focal}. By focusing on the minority and hard-to-classify ARB-prone class, Focal Loss significantly improves the model’s performance on extremely imbalanced data.

To reduce cross-domain distribution discrepancies, ARFT-Transformer applies the same tokenization and Transformer layer processing to the target domain as it does to the source domain, and computes the MMD loss between the source domain features (including [CLS]) and the target domain features. Given the distribution of source domain $P_s$ and target domain distribution ${P_t}$,  MMD is defined as the difference between their mean embeddings in the high-dimensional feature space:
\begin{equation}
\mathrm{MMD}(\mathcal{F}, P_s, P_t) = \left\| \mathbb{E}_{x_s \sim P_s} [\phi(X_s)] - \mathbb{E}_{x_t \sim P_t} [\phi(X_t)] \right\|_{\mathcal{H}},  \label{eq:mmd_def}
\end{equation}
where $\phi(X_s)$ and $\phi(X_t)$ denote the output of the final Transformer layer of FT-Transformer. $\mathcal{H}$ represents the Reproducing Kernel Hilbert Space (RKHS, a high-dimensional space defined by a kernel function) where MMD measures the distribution discrepancy. In this paper, we use the Radial Basis Function (RBF) kernel, which is defined as:  
\begin{equation}
k\left ( x_s,x_t \right ) = \mathrm{exp}\left ( -\frac{\left \| x_s-x_t \right \|^2 }{2\sigma^2}  \right ) ,
\label{eq:kernal}
\end{equation}
where RBF kernel as the specific implementation of $k\left ( x_s,x_t \right ) $, and $\left \| x_s-x_t \right \| ^2$ represents the Euclidean distance between the feature vectors $x_s$ and $x_t$. The value of $\sigma$ is determined using a default heuristic method, specifically set as a function of the median pairwise distances between samples, to balance local and global distribution characteristics.

During training, the model employs the Stochastic Gradient Descent (SGD) algorithm to optimize the model parameters, including momentum to accelerate convergence. We introduce a regularization of the momentum coefficient $\mu$ and $L2$  in the optimizer to reduce the risk of overfitting. ARFT-Transformer adopts a dynamically decaying learning rate, enabling rapid convergence in the early stages and gradually reducing the learning rate later to ensure better learning of ARB features in imbalanced datasets. Additionally, $10^{-3}$ is used as the initial learning rate.

\section{Experimental Setup}\label{sec:preparation}
This section presents the experimental setup in detail, including research questions, datasets, training and testing dataset setting and model evaluation criteria.

\subsection{Research questions}

In this paper, we aim to design experiments to address the following research questions:
\begin{itemize}
\item RQ1: What is the performance of ARFT-Transformer in single-source setting?
\item RQ2: How does ARFT-Transformer perform in multi-source setting?
\item RQ3: In the ARFT-Transformer framework, do the attention mechanism and the Focal Loss function influence the model's predictive performance?
\item RQ4: How does ARFT-Transformer perform compared with traditional feature selection methods?
\end{itemize}

\subsection{Datasets}\label{AA}
We select three open-source datasets to validate the effectiveness of our proposed approach: MySQL, Linux, and Apache HTTPD Server (HTTPD). These datasets are representative in the field of ARB prediction and have been widely adopted in prior studies \cite{qin2018studying,wan2019supervised,xu2020cross,li2021cross,xie2023cross}
. Detailed information about the software used in this paper is presented in Table~\ref{tab:ARB}, including the programming language, versions, and the number and proportion of ARB-prone files. 

The ARB labels used in this work were collected through manual analysis in previous studies \cite{cotroneo2013predicting,qin2018studying}. Each closed and fixed bug report was examined in detail, including its descriptions, developer discussions, and patch files. A bug was classified as an ARB if it exhibited accumulating internal error states or if its activation or propagation behavior depended on the total system runtime. Source code files that were fixed to remove ARBs were labeled as ARB-prone (positive), whereas all remaining files were labeled as ARB-free (negative). As shown in the table, the proportion of ARB-prone files across all datasets is relatively small, which aligns with the sparsity characteristic of ARBs.

\begin{table*}[!t]
\renewcommand{\arraystretch}{1.2}
\caption{\textsc{ARB Information}}
\vspace{-2mm}
\setlength{\tabcolsep}{4pt}  
\centering
\begin{tabular}{c c c c c c c c c}
\toprule
Projects & Language & Version & Components & Time Period & ARBs & Files & ARB-prone Files & ARB-prone Files\% \\
\midrule
Linux & C & 2.6 & \makecell[c]{Network Drivers \\ SCSI Drivers \\ EXT3 Filesystem \\ Networking/IPV4} & 2003.12--2011.5 & 20 & 3400 & 20 & 0.59\% \\
\hline
MySQL & C++/C & 5.1 & \makecell[c]{InnoDB Storage Engine \\ Replication \\ Optimizer} & 2003.8--2011.12 & 16 & 730 & 41 & 5.62\% \\
\hline
HTTPD & C & 2.0 & \makecell[c]{all} & 2002.5--2016.12 & 23 & 803 & 17 & 2.12\% \\
\bottomrule
\end{tabular}
\label{tab:ARB}
\vspace{-1mm}
\end{table*}

\begin{table*}[!t]
\renewcommand{\arraystretch}{1.2}
\caption{\textsc{Summary of the Metrics}}
\vspace{-2mm}
\hspace*{-4mm} 
\setlength{\tabcolsep}{2pt}
\centering  
\renewcommand{\arraystretch}{1.2}
\begin{tabular}{c|l}
\hline
\textit{Type} & \textit{Metrics}   \\
\hline
Program size  
& \begin{tabular}[l]
{@{}l@{}}
AltAvgLineBlank, AltAvgLineCode, AltAvgLineComment, AltCountLineBlank, AltCountLineCode,\\  AltCountLineComment, AvgLine, AvgLineBlank, AvgLineCode, AvgLineComment, \\CountDeclClass,  CountDeclFunction, CountLine, CountLineBlank, CountLineCode,\\ CountLineCodeDecl, CountLineCodeExe,  CountLineComment, CountLineInactive, \\CountLinePreprocessor, CountSemicolon, CountStmt, CountStmtDecl, CountStmtEmpty,\\ CountStmtExe, RatioCommentToCode\end{tabular} \\
\hline
McCabe’s complexity          
& \begin{tabular}[l]{@{}l@{}}
AvgCyclomatic, AvgCyclomaticModified, AvgCyclomaticStrict, AvgEssential, MaxCyclomatic,\\ MaxCyclomaticModified, MaxCyclomaticStrict, SumCyclomatic, SumCyclomaticModified,\\  SumCyclomaticStrict, SumEssential\end{tabular}     \\
\hline
Halstead metrics             
& \begin{tabular}[c]{@{}c@{}}Program Volume, Program Length, Program Vocabulary, Program Difficulty, Effort, N1, N2, n1, n2\end{tabular} \\
\hline
Aging-Related Metrics (ARMs) 
& \begin{tabular}[c]{@{}c@{}}AllocOps, DeallocOps, DerefSet, DerefUse,  UniqueDerefSet, UniqueDerefUse\end{tabular} \\
\hline
\end{tabular}
\label{tab:metrics}
\vspace{-2mm}
\end{table*}

The software metrics are all static code metrics used in prior work \cite{qin2018studying,wan2019supervised,xu2020cross,li2021cross,xie2023cross}, including program size metrics, McCabe complexity metrics, Halstead metrics, and Aging-related metrics as shown in Table~\ref{tab:metrics}. Among these, Aging-related metrics are specifically designed for ARBs. $AllocOps$, or “Allocation Operations,” refers to the number of memory allocation primitives (e.g., \textit{malloc} or \textit{new}) called in a file. Similarly, $DeallocOps$, or “Deallocation Operations,” indicates the number of memory deallocation primitives (e.g., \textit{free} or \textit{delete}) called. $DerefSet$ and $DerefUse$ represent the total number of times pointer variables are dereferenced in a file, where $DerefSet$ counts dereferences used for assigning values, and $DerefUse$ counts those used for reading values \cite{cotroneo2013predicting}. $UniqueDerefSet$ and $UniqueDerefUse$ represent the number of distinct pointer variables dereferenced in write and read operations, respectively. Unlike total counts, these metrics reflect the diversity of pointer usage in the code. Program size metrics \cite{akiyama1971example} (e.g., \textit{AltAvgLineBlank}, \textit{AltAvgLineCode}) are related to lines of code, statements, comments and declarations, providing a basic estimate of software complexity \cite{cotroneo2013predicting}. McCabe complexity metrics \cite{mccabe1976complexity} (e.g., \textit{AvgCyclomatic}, \textit{SumEssential}) are derived from the control flow graph of a program, measuring complexity through the nodes (code blocks) and edges (branches). Halstead metrics \cite{halstead1977elements} (e.g., \textit{Program Volume}, \textit{Program Length}) are defined based on the number of operators and operands in the program.

\subsection{Training and Testing Dataset Setting}

To evaluate the predictive performance of the proposed method, we design two types of experiments. 

\subsubsection{Single-source type}
In the single-source type, we perform cross-project ARB prediction by using one of MySQL, Linux, and HTTPD as target dataset and another as source dataset, designing six experiments: L$\Rightarrow$M, H$\Rightarrow$M, H$\Rightarrow$L, M$\Rightarrow$L, M$\Rightarrow$H, and L$\Rightarrow$H where L, M, H are short for Linux, MySQL and HTTPD respectively, the left and right item of $\Rightarrow$ denotes source and target project. For example, L$\Rightarrow$M refers to the prediction scenario using Linux as source project and MySQL as target project.

\subsubsection{Multi-source type}
In the multi-source type, we conduct prediction by using one project as target set and the remaining two datasets combined as source set, designing three experiments: LH$\Rightarrow$M, MH$\Rightarrow$L, and ML$\Rightarrow$H. 

\begin{table}[!t]
\renewcommand{\arraystretch}{1.2}
\caption{\textsc{Confusion Matrix}}
\vspace{-6mm}
\begin{center}
\begin{tabular}{c|c|c}
\hline
\multirow{2}{*}{\textbf{Actual Class}} & \multicolumn{2}{c}{\textbf{Predicted Class}} \\
\cline{2-3} 
 & \textbf{\textit{ARB-prone}} & \textbf{\textit{ARB-free}} \\
\hline
ARB-prone & TP & FN \\
\hline
ARB-free & FP & TN \\
\hline
\end{tabular}
\label{tab:matrix}
\end{center}
\vspace{-2mm}
\end{table}

\subsection{Model Evaluation Criteria}
\label{subsec: model_evaluation_criteria}

Given the highly imbalanced nature of the ARB prediction task (extremely low proportion of ARB-prone class samples), we use evaluation measures that are suitable for this task and have been widely used in prior studies \cite{qin2018studying,wan2019supervised,xu2020cross,li2021cross,xie2023cross}, including Probability of Detection ($PD$), Probability of False Alarms ($PF$), and Balance ($Bal$). All these metrics are derived from the confusion matrix as shown in Table~\ref{tab:matrix}. The components of the metric are as follows: True Positive ($TP$) represents the number of files correctly identified by the model as ARB-prone; False Negative ($FN$) denotes the number of files that are actually ARB-prone but are incorrectly predicted by the model as ARB-free; False Positive ($FP$) represents the number of ARB-free files incorrectly predicted by the model as ARB-prone; True Negative ($TN$) indicates the number of files correctly predicted by the model as ARB-free.

The definition of the evaluation measures are as follows:
$PD$ focuses on the detection performance of ARB-prone class samples and is unaffected by the number of ARB-free samples, thus providing stable evaluation results even in scenarios with sparse ARB-prone classes. A higher $PD$ value indicates a better model performance.
\begin{equation}
PD = \frac{TP}{TP + FN}
\end{equation}

$PF$ focuses on the prediction performance for ARB-free samples, measuring how effectively the model identifies ARB-free files. A lower $PF$ value indicates a better model performance, as it reflects fewer false positives and more accurate detection of the ARB-free class.
\begin{equation}
PF = \frac{FP}{FP + TN}
\end{equation}

$Bal$ is directly derived from the values of $PD$ and $PF$, considering both the ARB-prone class detection ($PD$) and ARB-free class misclassification ($PF$) to reflect the model’s trade-off. Compared to using $PD$ or $PF$ individually, $Bal$ better captures the overall performance. A higher $Bal$ value indicates a better comprehensive model performance.

\begin{equation}
Bal=1-\frac{\sqrt{\left ( 0-PF\right)^2 +\left ( 1-PD\right)^2}}{\sqrt{2} }
\end{equation}

\begin{table}[!t]
\renewcommand{\arraystretch}{1.2}
\caption{\textsc{Prediction Performance in single-source setting}}
\vspace{-2mm}
\hspace{-5mm}
\setlength{\abovecaptionskip}{0pt}  \label{table:single-source}
\begin{tabular}{p{6mm} c c c c c c p{6mm}}
\hline
{Group} & {H$\Rightarrow$L} & {M$\Rightarrow$L} & {H$\Rightarrow$M} & {L$\Rightarrow$M} & {M$\Rightarrow$H} & {L$\Rightarrow$H} & {Avg.} \\
\hline
PD & \textbf{0.992} & 0.981 & 0.894 & 0.867 & 0.866 & 0.863 & 0.912 \\
PF & 0.287 & 0.292 & 0.234 & 0.167 & 0.113 & \textbf{0.092} & 0.197 \\
Bal & 0.794 & 0.789 & 0.777 & 0.807 & 0.846 & \textbf{0.856} & 0.812 \\
\hline
\end{tabular}
\vspace{-4mm}
\end{table}

\section{Experimental Results}
\label{sec:results}
In this section, we will address three research questions proposed in Section \ref{introduction}, presenting the experimental results and the performance of our model.

\subsection{RQ1: What is the performance of ARFT-Transformer in single-source setting?}

The experimental results of single-source type is shown in Table \ref{table:single-source}, in which `Avg.' column denotes the average prediction performance among experimental groups and the group having the best performance under each evaluation measure is marked in boldface. From the table, we can see that all $Bal$ values exceed 0.777, with the average value of 0.812 and highest value achieving 0.856 when the model is trained on Linux and tested on HTTPD. The good performance of $Bal$ stems from the relatively high value of $PD$ and low value of $PF$ with average value of 0.912 and and 0.197 respectively.

To further exhibit the performance of ARFT-Transformer, we also compare its performance in terms of comprehensive evaluation measure $Bal$ with six state-of-the-art methods (a detailed introduction is provided in Section \ref{sec:relatedwork}), and list the results in Table \ref{tab:six-method}. A detailed introduction to the six state-of-the-art methods is provided in Section 8 (Related Work). To perform a fair comparison, we compare to the methods experimented with the same software metrics as listed in this paper. Note that the values of SRLA, JDA-ISDA, JPKS and KDK shown in the table correspond to the best performance of the compared methods. Since the JDA-ISDA, JPKS, KDK and BISP methods do not experiment with HTTPD project, we show their performance with Linux and MySQL. In the table, `Impv.' shows the improvement of ARFT-Transformer ($b$) when compared with baseline methods ($a$) and is calculated as $(b-a)/a\%$. The values indicating a better performance of ARTF-Transformer are in bold.

\begin{table*}[!t]
\renewcommand{\arraystretch}{1.2}
\caption{\textsc{The Comparison between ARFT-Transformer and State-of-the-art methods in single-source case}}
\setlength{\abovecaptionskip}{0pt}
\vspace{-5pt}
\centering
\setlength{\tabcolsep}{10pt} 
\begin{tabular}{c c c c c c c c}
\hline
{Group} & {H$\Rightarrow$L} & {M$\Rightarrow$L} & {H$\Rightarrow$M} & {L$\Rightarrow$M} & {M$\Rightarrow$H} & {L$\Rightarrow$H} & {Avg.} \\
\hline
TLAP \cite{qin2018studying} & 0.721 & 0.754 & 0.745 & 0.744 & 0.694 & 0.689 & 0.724 \\
Impv. & \textbf{16.125\%} & \textbf{6.641\%} & \textbf{4.295\%} & \textbf{8.468\%} & \textbf{21.902\%} & \textbf{24.238\%} & \textbf{12.278\%} \\
\hline
TLAP(C+N) \cite{qin2023predicting} & 0.773 & 0.766 & 0.601 & 0.623 & 0.666 & 0.692 & 0.686 \\
Impv. & \textbf{2.717\%} & \textbf{3.002\%} & \textbf{29.285\%} & \textbf{29.535\%} & \textbf{27.027\%} & \textbf{23.699\%} & \textbf{19.211\%} \\[0.5ex]
\hline
SRLA \cite{wan2019supervised}& 0.725 & 0.791 & 0.731 & 0.743 & 0.708 & 0.704 & 0.733 \\
Impv. & \textbf{9.517\%} & -0.253\%& \textbf{6.293\%} & \textbf{8.614\%} & \textbf{19.492\%} & \textbf{21.591\%} & \textbf{10.876\%} \\[0.5ex]
\hline
JDA-ISDA \cite{xu2020cross}& \multicolumn{1}{c}{--} & 0.791 & \multicolumn{1}{c}{--} & 0.772 & \multicolumn{1}{c}{--} & \multicolumn{1}{c}{--} & 0.781 \\
Impv. & \multicolumn{1}{c}{--} & -0.253\% & \multicolumn{1}{c}{--} & \textbf{4.534\%} & \multicolumn{1}{c}{--} & \multicolumn{1}{c}{--} & \textbf{2.141\%} \\[0.5ex]
\hline
JPKS \cite{li2021cross}& \multicolumn{1}{c}{--} & 0.793 & \multicolumn{1}{c}{--} & 0.774 & \multicolumn{1}{c}{--} & \multicolumn{1}{c}{--} & 0.781 \\
Impv. & \multicolumn{1}{c}{--} & -0.507\% & \multicolumn{1}{c}{--} & \textbf{4.264\%} & \multicolumn{1}{c}{--} & \multicolumn{1}{c}{--} & \textbf{1.879\%} \\[0.5ex]
\hline
KDK \cite{xie2023cross}& \multicolumn{1}{c}{--} & 0.782 & \multicolumn{1}{c}{--} & 0.787 & \multicolumn{1}{c}{--} & \multicolumn{1}{c}{--} & 0.784 \\
Impv. & \multicolumn{1}{c}{--} & \textbf{0.895\%} & \multicolumn{1}{c}{--} & \textbf{2.541\%} & \multicolumn{1}{c}{--} & \multicolumn{1}{c}{--} & \textbf{1.718\%} \\[0.5ex]
\hline
BISP \cite{xu2025cross}& \multicolumn{1}{c}{--} & 0.772 & \multicolumn{1}{c}{--} & 0.759 & \multicolumn{1}{c}{--} & \multicolumn{1}{c}{--} & 0.766 \\
Impv. & \multicolumn{1}{c}{--} & \textbf{2.202\%} & \multicolumn{1}{c}{--} & \textbf{6.324\%} & \multicolumn{1}{c}{--} & \multicolumn{1}{c}{--} & \textbf{4.263\%} \\[0.5ex]
\hline
\end{tabular}
\label{tab:six-method}
\vspace{-1pt}
\end{table*}

From the table, it is obvious that ARTF-Transformer exhibits a better performance on average, with an average improvement of 12.278\%, 19.211\%, 10.876\%, 2.141\%, 1.879\%, 1.718\% and 4.263\% when compared with TLAP, TLAP (C+N), SRLA, JDA-ISDA, JPKS, KDK and BISP respectively. In details, ARTF-Transformer demonstrates superiority in all the single-source groups in comparison with TLAP, TLAP (C+N) and BISP, with the largest improvement of 24.238\%, 29.535\%, and 6.324\% respectively. Although one group (i.e., M$\Rightarrow$L) shows inferior performance in SRLA, JDA-ISDA and JPKS, the decrease percentage is not large, with values less than 0.6\%\textemdash at least one order of magnitude smaller than the improvement.  

Based on the experimental results, we can conclude that ARFT-Transformer shows a good prediction performance in single-source setting, achieving a better average performance than the state-of-the-art methods.

\begin{table}[!t]
\renewcommand{\arraystretch}{1.2}
\caption{\textsc{Prediction Performance in multi-source case}}
\setlength{\abovecaptionskip}{0pt} \label{table:multi-source}
\vspace{-5mm}
\begin{center}
\begin{tabular}{c c c c c}
\hline
{Group} & {MH$\Rightarrow$L} & {LH$\Rightarrow$M} & {LM$\Rightarrow$H} & {Avg.} \\
\hline
PD & \textbf{0.988} & 0.888 & 0.854 & 0.910 \\
PF & 0.224 & 0.213 & \textbf{0.081} & 0.172 \\
Bal & 0.836 & 0.785 & \textbf{0.861} & 0.827\\
\hline
\end{tabular}
\end{center}
\vspace{-1mm}
\end{table}

\subsection{RQ2: How does ARFT-Transformer perform in multi-source setting?}
The experimental results of multi-source case is shown in Table \ref{table:multi-source}. It can be observed from the table that two out of three groups show $Bal$ values larger than 0.8, and the rest reaches 0.785. We can also notice that the average performance of multi-source case is larger than the single-source case, with the values of 0.827 and 0.812 in multi-source and single-source case respectively. We wonder whether utilizing multi-source projects has an advantage over single-source project when dealing with the same target project and exhibit the comparison in Table \ref{table:comparison-pairwise-multi}. From the table, we can get the multi-source case behaves best in two scenarios in which Linux and HTTPD serve as the target project respectively. When MySQL serves as the target project, the LH$\Rightarrow$M behaves better than H$\Rightarrow$M but less than L$\Rightarrow$M. This phenomenon echoes the conclusion in \cite{qin2018studying} in which multi-source prediction shows an improvement over at least one single-source prediction for tackling the prediction task of same target project.

\begin{table}[!t]
\renewcommand{\arraystretch}{1.2}
\caption{\textsc{The Comparison Between multi-source and single-source case with the same target project}}
\setlength{\abovecaptionskip}{0pt} \label{table:comparison-pairwise-multi}
 \vspace{-5mm}
\begin{center}
\begin{tabular}{c c | c c | c c}
\hline
Group& Bal& Group& Bal & Group&Bal\\
\hline
H$\Rightarrow$L& 0.794& H$\Rightarrow$M& 0.777& M$\Rightarrow$H&0.846\\
M$\Rightarrow$L& 0.789& L$\Rightarrow$M& \textbf{0.807}& L$\Rightarrow$H&0.856\\
MH$\Rightarrow$L & \textbf{0.836} & LH$\Rightarrow$M & 0.785 & LM$\Rightarrow$H &\textbf{0.861} \\
\hline
\end{tabular}
\end{center}
\vspace{-8pt}
\end{table}

Similar to single-source prediction case, we also compare the performance of ARFT-Transformer with previous methods in multi-source case. Since only TLAP conducts multi-source cross-project ARB prediction, we show the improvement of ARFT-Transformer over TLAP in Table \ref{table:comparison-multi}. It can be derived from the table that ARFT-Transformer has a better performance over TLAP in three groups, with an improvement of 10.436\%, 5.795\% and 19.916\% respectively. The average improvement is 11.953\%.

\begin{table}[!t]
\renewcommand{\arraystretch}{1.2}
\caption{\textsc{The Comparison between ARFT-Transformer and State-of-the-art methods in multi-source case}}
\setlength{\abovecaptionskip}{0pt} \label{table:comparison-multi}
 \vspace{-5mm}
\begin{center}
\begin{tabular}{c c c c c}
\hline
Group& MH$\Rightarrow$L& LH$\Rightarrow$M& LM$\Rightarrow$H &Avg.\\
\hline
TLAP \cite{qin2018studying}& 0.757& 0.742& 0.718 &0.739\\
Impv.& \textbf{10.436\%}& \textbf{5.795\%}& \textbf{19.916\%} &\textbf{11.953\%}\\
\hline
\end{tabular}
\end{center}
\vspace{-10pt}
\end{table}

To sum up, the experimental results indicate that ARFT-Transformer performs well in multi-source prediction setting, exhibiting a significant average superiority over state-of-the-art methods. In addition, the multi-source prediction has a better performance than single-source, demonstrating the usefulness of combining multiple source information during model training.

\begin{table*}[htbp]
\renewcommand{\arraystretch}{1.2}
\caption{\textsc{The Results of Ablation Experiment }}
\setlength{\abovecaptionskip}{0pt}
\setlength{\belowcaptionskip}{0pt}
\vspace{-1pt}
\centering
\setlength{\tabcolsep}{10pt} 
\begin{tabular}{c c c c c c c c}
\hline
{Group} & {H$\Rightarrow$L} & {M$\Rightarrow$L} & {H$\Rightarrow$M} & {L$\Rightarrow$M} & {M$\Rightarrow$H} & {L$\Rightarrow$H} & {Avg.} \\
\hline
Baseline & 0.603 & 0.541 & 0.446 & 0.415 & 0.836 & 0.347 & 0.531 \\
\hline
Baseline+Focal & 0.541 & 0.727 & 0.508 & 0.415 & 0.836 & 0.456 & 0.580 \\
Impv. & -10.282\% & \textbf{34.381\%} & \textbf{13.901\%} & 0.000\% & 0.000\% & \textbf{31.412\%} & \textbf{11.569\%} \\
\hline
Baseline+attent & 0.752 & 0.778 & 0.700 & 0.781 & 0.842 & 0.752 & 0.767 \\
Impv. & \textbf{24.710\%} & \textbf{43.810\%} & \textbf{56.951\%} & \textbf{88.193\%} & \textbf{0.718\%} & \textbf{116.715\%} & \textbf{55.183\%} \\
\hline
Baseline+attent+Focal & 0.794 & 0.789 & 0.777 & 0.807 & 0.846 & 0.856 & 0.812 \\
Impv. & \textbf{31.675\%} & \textbf{43.810\%} & \textbf{74.215\%} & \textbf{94.458\%} & \textbf{1.196\%} & \textbf{146.686\%} & \textbf{65.340\%} \\
\hline
\end{tabular}
\label{tab:ablation-experiment}
\vspace{-1pt}
\end{table*}

\subsection{RQ3: In the ARFT-Transformer framework, do the attention mechanism and the Focal Loss function influence the model's predictive performance?}
Since ARFT-Transformer incorporates both an attention mechanism and Focal Loss as shown in Figure~\ref{fig:model}, it is necessary to assess the individual contribution of each component. In this section, we conduct a series of ablation studies under three settings: (1) incorporating only the attention mechanism, (2) applying only Focal Loss, and (3) integrating both components (i.e., ARFT-Transformer).

The main part of ARFT-Transformer mainly includes tokenizer + multi-head attention mechanism + Focal Loss + MMD Loss. Comparing to the framework of ARFT-Transformer, the baseline model is composed of tokenizer + feedforward neural network (to replace the original multi-head attention mechanism) + Cross-entropy Loss (to replace the original Focal Loss) + MMD loss. Each neural network layer contains a feedforward neural network composed of two Linear layers, a ReGLU activation layer, and a Dropout layer. In the first setting, we apply a multi-head attention mechanism before the feed-forward neuron network of baseline, referred to as “baseline+attent”. In the second setting, we replace the cross-entropy loss in baseline with a Focal Loss incorporating a modulation factor, denoted as “baseline+Focal”. The setting that includes both modifications is called “baseline+attent+Focal” (i.e., ARFT-Transformer). Experiments are conducted across six pair-wise groups, where ‘Impv.’ represents the percentage improvement in the $Bal$ value compared to the baseline. Values that outperform the baseline are highlighted in bold.

The ablation study results are shown in Table~\ref{tab:ablation-experiment}. From the table, we can observe that when cross-entropy loss is replaced by Focal Loss, the $Bal$ values of the six pair-wise groups are improved by -10.282\%, 34.381\%, 13.901\%, 0.000\%, 0.000\%, and 31.412\%, with three groups showing better performance and two groups showing no change, resulting in an average improvement of 11.569\%. When multi-head attention is applied, the $Bal$ values of all six pair-wise experiments perform better, with the highest improvement reaching 116.715\% in the L$\Rightarrow$H scenario, and the average improvement of 55.183\%. It indicates that both Focal Loss and multi-head attention have a positive impact on the model in general, and adding multi-head attention mechanism has a larger improvement than the Focal Loss in each experimental group.

We also observe that when both Focal Loss and multi-head attention are applied together, the improvements in the six pair-wise groups achieve 31.675\%, 43.810\%, 74.215\%, 94.458\%, 1.196\%, and 146.686\%, respectively, with an average improvement of 65.340\%, demonstrating strong performance benefits. All six groups show performance improvements, with the lowest improvement of 1.196\% in M$\Rightarrow$H and the highest improvement of 146.686\% in L$\Rightarrow$H. When focusing on each group, we can find that combining multi-head attention and Focal Loss has a larger or equal improvement than using one of them alone, demonstrating the indispensability of two components.

In conclusion, ablation studies demonstrate that both multi-head attention and Focal Loss have a significant impact on the model’s evaluation performance, and the best results are achieved when they are combined together.

\subsection{RQ4: How does ARFT-Transformer perform compared with traditional feature selection methods?}

To further demonstrate the performance of ARFT-Transformer, we compare it with five traditional feature selection methods (OneR, Information Gain, Gain Ratio, RELEIF and Symmetric Uncertainty) which also addressed feature correlation among metrics in previous ARB prediction study \cite{kumar2018feature}. When using OneR, we find that all metrics are deleted, so its results are not considered in the comparison. To perform a fair comparison, we compare to the methods experimented with the same software metrics as listed in this paper. Since ARFT-Transformer uses a linear layer as the classifier and applies focal loss to monitor the training process, we replace the FT-Transformer in ARFT-Transformer with these feature selection methods respectively and use the same linear classifier with focal loss.

Table \ref{tab:feature-selection} shows the experimental results. In the table, the `Avg.' column denotes the average prediction performance among six experimental groups, and `Impv.' shows the improvement of ARFT-Transformer over each feature selection method. The `Impv.' values achieving a better performances are highlighted in bold. From the table, we can get that ARFT-Transformer shows higher average $Bal$ values than Information Gain, Gain Ratio, RELEIF, and Symmetric Uncertainty, with average improvements of 18.380\%, 100.078\%, 34.171\%, and 176.963\%, respectively. In details, ARFT-Transformer demonstrates superior performance in all the six groups. When trained on HTTPD and tested on MySQL, ARFT-Transformer achieves the lowest $Bal$ value of 0.777, but it is still higher than the best performance achieved using Information Gain, with an improvement of 3.187\%. 

Based on the experiment results, we can conclude that ARFT-Transformer achieves better performance than traditional feature selection methods in single-source setting demonstrating its ability to capture metric dependencies and extract crucial information.

\begin{figure*}[htbp]
    \centering
    \subfloat[Impact of the number of Heads in Multi-head Attention.]{%
        \includegraphics[width=0.45\textwidth]{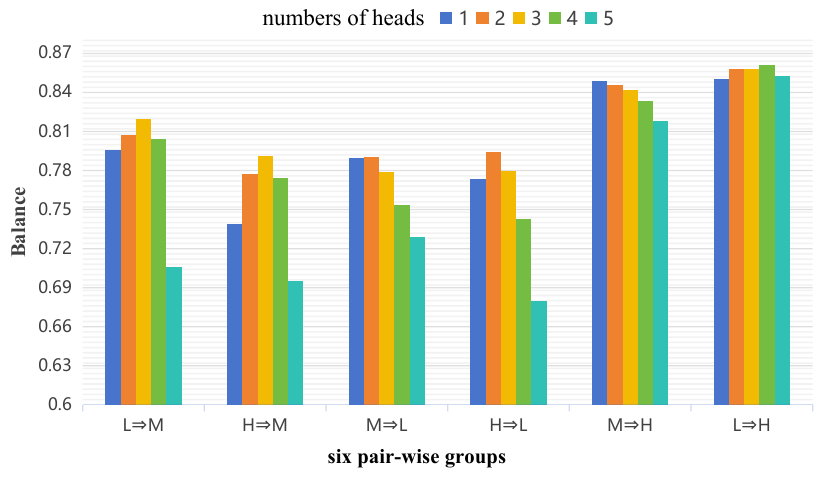}%
        \label{fig:n_heads}
    }
    \hspace{2mm}
    \subfloat[Impact of the modulating Factor of Focal Loss]{%
        \includegraphics[width=0.45\textwidth]{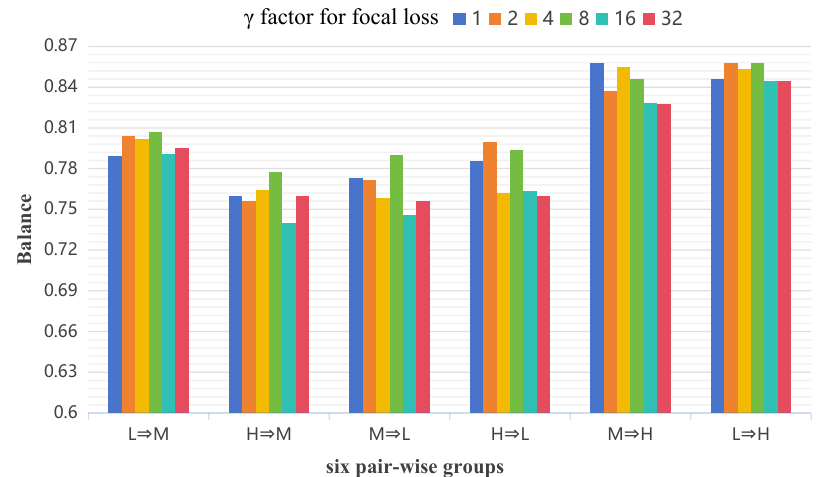}%
        \label{fig:gamma}
    }
    \caption{Influence of attention heads and focal loss modulation factor on model performance.}
    \label{fig:combined}
    
\vspace{-1pt}
\end{figure*}
\begin{table*}[!ht] 
\renewcommand{\arraystretch}{1.2} 
\setlength{\tabcolsep}{8pt} 
\centering
\caption{\textsc{The Comparison between ARFT-Transformer and feature selection methods in single-source case}}
\vspace{2pt} 
\begin{tabular}{c c c c c c c c}
\hline
{Group} & {H$\Rightarrow$L} & {M$\Rightarrow$L} & {H$\Rightarrow$M} & {L$\Rightarrow$M} & {M$\Rightarrow$H} & {L$\Rightarrow$H} & {Avg.} \\
\hline
ARFT & 0.794 & 0.789 & 0.777 & 0.807 & 0.846 & 0.856 & 0.812 \\
\hline
Information Gain & 0.595 & 0.700 & 0.753 & 0.751 & 0.697 & 0.648 & 0.691 \\
Impv. & \textbf{33.445\%} & \textbf{12.714\%} & \textbf{3.187\%} & \textbf{7.457\%} & \textbf{21.377\%} & \textbf{32.099\%} & \textbf{18.380\%} \\
\hline
Gain Ratio & 0.333 & 0.317 & 0.390 & 0.326 & 0.675 & 0.607 & 0.441 \\
Impv. & \textbf{138.438\%} & \textbf{148.896\%} & \textbf{99.231\%} & \textbf{147.546\%} & \textbf{25.333\%} & \textbf{41.021\%} & \textbf{100.078\%} \\
\hline
RELEIF & 0.648 & 0.668 & 0.681 & 0.692 & 0.434 & 0.617 & 0.623 \\
Impv. & \textbf{22.531\%} & \textbf{18.114\%} & \textbf{14.097\%} & \textbf{16.618\%} & \textbf{94.931\%} & \textbf{38.736\%} & \textbf{34.171\%} \\
\hline
Symmetric Uncertainty & 0.293 & 0.293 & 0.293 & 0.293 & 0.293 & 0.293 & 0.293 \\
Impv. & \textbf{170.990\%} & \textbf{169.283\%} & \textbf{165.188\%} & \textbf{175.427\%} & \textbf{188.737\%} & \textbf{192.150\%} & \textbf{176.963\%} \\
\hline
\end{tabular}
\vspace{2pt} 
\label{tab:feature-selection}
\end{table*}

\section{Sensitivity Analysis}
\label{sec:discussion}
In this section, we would like to discuss the impact of the choice of hyperparameters on the performance of ARFT-Transformer. It has two key hyperparameters: the number of attention heads $m$ in FT-Transformer, and the modulating factor $\gamma$ in Focal Loss. We vary above two hyperparameters and measure the change in prediction performance on six pair-wise groups in the single-source setting.

\subsection{Impact of the number attention heads in FT-Transformer}
To study the influence of attention head number $m$ on the prediction performance of ARFT-Transformer, we set the $\gamma$ parameter value as 2 and change the the number of attention heads as 1, 2, 4, 8, 16, and 32 respectively. Figure~\ref{fig:n_heads} shows the change of $Bal$ along with the variation of head number, with the horizontal axis representing six pair-wise groups and the vertical axis indicating the performance in terms of the $Bal$ metric. Each color in the legend represents the number of attention heads. From the figure, we can observe that a higher number of heads not always contributes to a better cross-project ARB prediction ability, the performance drops with too many heads. Among the six experimental groups, the best $Bal$ values are achieved when the number of attention heads is 2, 8, 8, 8, 1, and 2, respectively. In three out of six groups, the best results are obtained when using eight attention heads, i.e., 0.80 in L$\Rightarrow$M, 0.77 in H$\Rightarrow$M and 0.78 in M$\Rightarrow$L. We also calculate the average performance across the six pair-wise scenarios in each hyperparameter value, getting 0.801, 0.804, 0.798, 0.811, 0.785, 0.790 respectively. From the result, we can observe that ARFT-Transformer achieves the best average performance when using 8 attention heads.

\subsection{Impact of modulating factor in Focal Loss}
To study the impact of modulating factor $\gamma$ in Focal Loss across six pair-wise groups, we keep the number of attention heads fixed at 8 and vary $\gamma$ from 1 to 5. As illustrated in Figure \ref{fig:gamma}, the horizontal axis represents six pair-wise groups and the vertical axis shows the $Bal$ performance. Each color in the legend represents the value of $\gamma$. From the figure, we observe that the best $Bal$ values for the six experimental groups are achieved when $\gamma$ is set to 2, 2, 3, 3, 1, and 4, respectively. Two of the six groups achieve peak performance with $\gamma$ setting to 2 or 3. In terms of average performance across the six pair-wise scenarios, $\gamma$ values of 2 and 3 achieve 0.812 and 0.811, respectively. This suggests that ARFT-Transformer delivers the best average predictive performance when $\gamma$ is 2. 

\section{Threats to validity}
\label{sec:threats_to_validity}
\subsection{Threats to construct validity}
Threats to construct validity refers to the degree to which the measurements used in a study accurately reflect the theoretical concepts being investigated, including the quality of data labeling and the appropriateness of evaluation measures. Following prior research \cite{qin2018studying,wan2019supervised,xu2020cross,li2021cross,xie2023cross}, bug reports in our dataset were manually labeled as either ARB-prone or ARB-free. According to the original paper, the annotation work was performed independently by two authors, and the issue reports were labeled into two types after reaching a consensus, thus reducing potential subjectivity or bias in the classification.

Additionally, in our work, we adopt $PD$, $PF$, and $Bal$ as evaluation measurements to assess the effectiveness of the proposed approach following previous studies \cite{qin2018studying,wan2019supervised,xu2020cross,li2021cross,xie2023cross}. In future work, we plan to incorporate additional evaluation indicators, such as the F1-score, to provide a more comprehensive assessment of model performance.

\subsection{Threats to Internal validity}
Threats to internal validity refer to internal factors that may affect the results. Although we do not perform extensive hyperparameter tuning and used default setting following the suggested value in \cite{gorishniy2021revisiting,lin2017focal}, we are aware that different hyperparameter settings could potentially influence model performance as discussed in Section \ref{sec:discussion}. 

\subsection{Threats to External validity}
Threats to external validity mainly focuses on the generalization of the obtained results. In this study, we validate the performance of ARFT-Transformer on three open-source and well-known projects  in ARB prediction. We have no idea the performance of the method on other projects, and we will explore the use of more datasets in subsequent work.

\section{Related work}\label{sec:relatedwork}
In this section, we review the related work on ARB prediction.

Since Cotroneo et al. \cite{cotroneo2010software} pointed out the relationship between size code metrics, Halstead code metrics, and software aging, and proposed six aging-related metrics combined with other static metrics for ARB prediction \cite{cotroneo2013predicting}, ARB prediction has gained significant attention. Some studies focused on within-project ARB prediction, where both training and testing datasets come from the same project with the same input distribution. Sharma et al. \cite{sharma2018analysis} evaluated basic machine learning methods, including SVM, Naive Bayes, Logistic Regression, KNN, MLP, and Decision Tree, using ensemble learning models to obtain the optimal classifier. Chouhan et al. \cite{chouhan2021generative} proposed a generative adversarial networks-based method for class imbalance. Zhang et al. \cite{zhang2023ifcm} used the Fuzzy C-means algorithm to improve prediction performance for ARB-prone samples by removing overlapping samples from the ARB-free samples. Zhang et al. \cite{zhang2024sgt} employed Graph Neural Networks with Code Property Graphs and ROS to mitigate class imbalance. Tian et al. \cite{tian2025towards} utilized Spiking Neural Networks to represent information and Bio-inspired, Diversity-aware Active Learning to label key samples, achieving good prediction results.

When sufficient training data are available, models achieve good prediction results. However, ARB datasets collection is challenging. To overcome the limited availability of training data for within-project ARB prediction, Qin et al. \cite{qin2015cross,qin2018studying} first combined TCA with ROS and proposed a transfer learning-based cross-project ARB prediction method (TLAP). However, cross-project ARB prediction faces challenges: class imbalance caused by the small proportion of ARB-prone samples and distribution discrepancy between source and target projects.

To address these challenges, Wan et al. \cite{wan2019supervised} explored feature reconstruction combined with label encoding of the source project, and proposed a cross-project ARB prediction method based on an autoencoder (SRLA). Xu et al. \cite{xu2020cross} focused on the issue of conditional distribution alignment and introduced a novel class imbalance handling strategy. They proposed JDA-ISDA, which integrated joint distribution adaptation with an enhanced sub-class discriminant analysis to improve distribution alignment and classification accuracy through subclass division. Li et al. \cite{li2021cross} considered both distribution adaptation and class imbalance by proposing JPKS, a method based on joint probability domain adaptation and k-means-based SMOTE to achieve more balanced and transferable feature representations. Xie et al. \cite{xie2023cross} proposed a method combining Kernel Principal Component Analysis (KPCA) and Double Marginalized Denoising Autoencoder (DMDA) to mitigate distribution discrepancy, while introducing K-means Clustering Cleaning Ensemble (KCE) to effectively address class imbalance. Additionally, Qin et al. \cite{qin2023predicting} proposed ARB metrics based on complex network theory, and analyzed its applicability and predictive performance in both within-project and cross-project ARB prediction scenarios. Zhao et al. \cite{zhao2025negcparbp} introduced the IK-hidden algorithm, which retains key features for constructing a Negative Database, and employed TLAP as the predictive model for ARB prediction. More recently, Xu et al. \cite{xu2025cross} combines balanced distribution adaptation (BDA), improved subclass discriminant analysis (ISDA), and self-paced ensemble under-sampling (SPE) to jointly address class imbalance and overlapping issues in cross-project ARB prediction. These studies provide alternative viewpoints and technical innovations that further enrich the research landscape of ARB prediction.

Although these studies have gradually improved the prediction effectiveness of cross-project ARB prediction, they all treat the metrics as independent and ignore the relationship among metrics. Furthermore, these studies simply employ resampling methods to increase the classifier’s attention to minority class, without distinguishing between easy-to-classify and hard-to-classify samples. Different from above studies, this paper models the relationships between metrics and applies Focal Loss to consider sample classification difficulty during model training.

\section{Conclusion and Future work}\label{sec:conclusion}
In this paper, we propose a cross-project ARB prediction approach named ARFT-Transformer by considering both the relationship among metrics with FT-Transformer, and dynamically down-weighting well-classified samples and focusing the model’s learning on hard samples with Focal Loss. To validate the effectiveness of the approach, we conduct single-source and multi-source experiments on three widely-used datasets, and compare their results with state-of-the-art methods. The experimental results show that the average performance of ARFT-Transformer outperforms the existing methods in both single-source and multi-source cases. In addition, ablation studies are performed to evaluate the impact of key components. 

In future work, we plan to further investigate the following aspects. First, we will explore Multi-source cross-project ARB prediction to further improve the ARB prediction performance. Second, we will consider more evaluation metrics to assess the model performance. Furthermore, we will consider datasets written in different languages.

\section*{Acknowledgements}
This work was supported by the National Natural Science Foundation of China [grant numbers 62403044 and 62372021], Natural Science Youth Foundation of Jiangsu Province, China [grant number BK20250443], Talent Fund of Beijing Jiaotong University [grant number 2025JBRC017], Laboratory Open Fund of Capital Normal University [grant number LIP2025S300].



\bibliographystyle{elsarticle-num} 
\bibliography{references}





\end{document}